\documentclass[useAMS,usenatbib]{mn2e}

\usepackage{multirow}
\usepackage{graphicx}
\usepackage{amsmath}

\def\apj{ApJ}
\def\apjl{ApJ}
\def\araa{ARA\&A}
\def\aap{A\&A} 
\def\apss{Ap\&SS}
\def\jcp{J.~Chem.~Phys.}
\def\jgr{J.~Geophys.~Res.}
\def\mnras{MNRAS}
\def\nat{Nature}
\def\prb{Phys.~Rev.~B}
\def\pccp{PCCP}
\def\ssr{Space~Sci.~Rev.}

\title[Thermal desorption characteristics of CO, O$_2$ and CO$_2$ on H$_2$O$_{(np)}$, H$_2$O$_{(c)}$ and SiO$_x$.]{Thermal desorption characteristics of CO, O$_2$ and CO$_2$ on non-porous water, crystalline water and silicate surfaces at sub-monolayer and multilayer coverages.}
\author[J. A. Noble et al.]{J.~A.~Noble$^{1,2}$, E.~Congiu$^{1}$, F.~Dulieu$^{1}$\thanks{E-mail: francois.dulieu@obspm.fr} and H.~J.~Fraser$^{2}$\\
$^{1}$Universit\'{e} de Cergy Pontoise, LERMA LAMAp, F-95000 Cergy Pontoise, France.\\
$^{2}$Department of Physics, Scottish Universities Physics Alliance (SUPA), University of Strathclyde, \\John Anderson Building, 107 Rottenrow East, Glasgow G4 ONG, Scotland, U.K.}
\begin{document}

\date{Accepted 2011 December 7.  Received 2011 December 6; in original form 2011 July 29}

\pagerange{\pageref{firstpage}--\pageref{lastpage}} \pubyear{2002}

\maketitle

\label{firstpage}

\begin{abstract}
The desorption characteristics of molecules on interstellar dust grains are important for modelling the behaviour of molecules in icy mantles and, critically, in describing the solid-gas interface. In this study, a series of laboratory experiments exploring the desorption of three small molecules from three astrophysically relevant surfaces are presented. The desorption of CO, O$_2$ and CO$_2$ at both sub-monolayer and multilayer coverages was investigated from non-porous water, crystalline water and silicate surfaces. Experimental data was modelled using the Polanyi-Wigner equation to produce a mathematical description of the desorption of each molecular species from each type of surface, uniquely describing both the monolayer and multilayer desorption in a single combined model. The implications of desorption behaviour over astrophysically relevant timescales are discussed.
\end{abstract}


\section{Introduction}

The interaction of molecular species with surfaces is of critical importance in astrophysical environments. Much interstellar chemistry occurs on or in the icy layers which cover dust grains in molecular clouds, the birthplace of stars and planets \citep{Fraser02}. These dust grains are believed to be composed of silicates and carbonaceous material \citep{Draine03}. The silicate is at least $\sim$~95~\% amorphous, as determined from the breadth of the 9.7~$\mu$m band \citep{Li2002}, and although its composition has not been determined precisely, is well approximated by amorphous olivine (MgFeSiO$_4$) \citep{Sofia01,Draine03}. The silicates are believed to be 'fluffy' in nature, with a large surface area upon which reactions can occur \citep{Mathis98}.

At the low temperatures in molecular clouds (T~$<$~20~K) atoms and molecules freeze out onto these silicate surfaces, forming the icy mantles. Amorphous solid water (ASW) is by far the largest component of this ice, with abundances of $\sim$~1~$\times$~10$^{-4}$ with respect to the total H column density \citep{Williams02}, equivalent to coverages of up to 100 monolayers (ML). Given that the extinction threshold for H$_2$O mantles is A$_V$~$\sim$~3.3 mag \citep{Whittet88}, it is reasonable to assume that between the dense pre-stellar cores (where dust grains are completely coated by this icy mantle) and the cloud edges (where competition between ice formation and photodesorption of H$_2$O yields a population of bare silicate grains \citep{Smith93}), there must be a region where icy surfaces and bare silicates co-exist.

As interstellar regions evolve, the icy grains are further processed, either by gentle heating or cyclic desorption-deposition events \citep{Visser09}, producing crystalline H$_2$O. This has been detected in various stellar and pre-planetary environments, including, for example, M giant stars \citep{Omont90}, Quaoar in the Kuiper Belt \citep{Jewitt04}, Trans Neptune Objects \citep{Merlin07}, comets \citep{Lisse06}, and in the outer disks around T Tauri stars \citep{Schegerer10}. 

As gas-grain modelling of interstellar environments becomes more sophisticated \citep{Wakelam10,Acharyya11,Cuppen11}, key questions impeding the full implementation of surface chemistry include: what effect does the underlying grain surface have on the desorption characteristics of key molecules, and to what extent is this desorption affected as we move from the multilayer to sub-monolayer coverage regime? A third issue is how to realise the transition between the multilayer and monolayer regimes in a gas-grain model without overloading the model with many more layers of complexity.

CO, O$_2$ and CO$_2$ are all interstellar molecules that potentially could populate bare interstellar grains at sub-monolayer coverages. On ice, only one previous study focussed on sub-monolayer coverages of CO, highlighting the spectroscopic rather than desorption characteristics of the porous ASW:CO system \citep{Collings05}. To date, no desorption studies of CO, O$_2$ or CO$_2$ have been made on a silicate surface, at either multilayer or sub-monolayer coverages.

CO is the second most abundant interstellar molecule \citep{Tielens05}, and is known to form in the gas phase, then freeze-out on to H$_2$O covered grains to form overlayers of pure CO ice, the layer thickness critically depending on gas density rather than grain temperature \citep{Pontoppidan03}.  Consequently, extensive temperature programmed desorption (TPD) studies have been made of multilayer CO coverages on various surfaces, including Au \citep{Acharyya07,Bisschop06,Collings04,Fuchs06}, H$_2$O \citep{Collings03a,Collings03b}, a meteorite sample \citep{Mautner06} and highly oriented pyrolytic graphite (HOPG) \citep{Ulbricht06}. 

It is widely accepted that both CO$_2$ and H$_2$O form on dust grains in molecular clouds, and recent studies suggest that at least some of the CO$_2$ and H$_2$O is formed concurrently \citep{Goumans08,Noble11,Ioppolo11}. Since CO is the key precursor to CO$_2$ formation, such mechanisms would require CO freeze-out onto bare grain surfaces long before CO ice, or even large quantities of H$_2$O ice, are detected. Although tenuous, spectroscopic observational and experimental evidence exists for such a freeze-out process \citep{Fraser05}, but an investigation of the desorption behaviour of CO at sub-monolayer coverages on ice and silicate surfaces is vital if we want to be certain that it can reside at an interstellar grain surface long enough to form more complex species. 

Similarly, O$_2$ is a potential precursor in H$_2$O formation \citep{Ioppolo08,Oba09}, and is also a species likely to form in the gas phase then freeze-out onto grain surfaces, rather than forming on the grain itself. However, as a homonuclear diatomic O$_2$ is infrared inactive, so its detection in interstellar ice, though occasionally claimed via a forbidden transition \citep{Elsila97} remains elusive. It is not clear whether this is related to the weak transition probability, lack of a significant O$_2$ population in the ice, or that the O$_2$ has rapidly reacted (upon adsorption) to exclusively form H$_2$O.  Nevertheless, the multilayer desorption behaviour of O$_2$ has been studied previously alongside CO and N$_2$ \citep{Acharyya07,Fuchs06}, as well as on Au, porous ASW \citep{Collings04} and TiO$_2$ \citep{Dohnalek06}. These studies show that the multilayer desorption characteristics of O$_2$ are very similar to CO, so it is interesting from both a chemical and astrophysical viewpoint to also investigate the desorption behaviour of O$_2$ at sub-monolayer coverages on ice and silicate surfaces.

The desorption characteristics of CO$_2$ have not been extensively studied on any surface, despite it being one of the most abundant solid phase molecular species in the interstellar medium. Recently the desorption characteristics of multilayer CO$_2$ were reported from porous ASW, and have been previously studied on HOPG \citep{Ulbricht06}, porous ASW and Au \citep{Collings04}. However, given that the observational \citep{Pontoppidan06} and experimental \citep{Noble11} evidence shows that a fraction of CO$_2$ ice must form concurrently with the water ice layer, some CO$_2$ molecules must populate both the bare silicate grains and the ice layers at sub-monolayer coverages. 

Here we present an experimental study of the desorption of CO, O$_2$ and CO$_2$ from three different surfaces; non-porous ASW, crystalline ice, and amorphous olivine-type silicate. As these are, for the first time, all undertaken in the same experimental set-up -- FORMOLISM \citep{Amiaud2006} -- we are able to investigate both the individual effect of each surface on the desorption characteristics, as well as determining whether the molecular composition or morphology of the surface is most relevant in determining the desorption behaviour of molecules. For each of the nine combinations of CO, O$_2$ and CO$_2$ on each surface, our study has encompassed both the multilayer and sub-monolayer regimes. These data are modelled to determine a simple analytical expression which accurately calculates both the sub-monolayer and the multilayer desorption energies. By changing the model to incorporate interstellar, rather than experimental, heating rates, we are able to address the key questions above, namely: what effect does the underlying surface have on the desorption characteristics of the molecules adsorbed there, and how is this desorption modified in the sub-monolayer coverage regime?

\section[]{Experimental}

The experiments were conducted using the FORMOLISM set-up (FORmation of MOLecules in the InterStellar Medium), described in detail elsewhere \citep{Amiaud2006,Lattelais11}. Briefly, the set-up consists of an ultra-high vacuum chamber (base pressure $\sim$~10$^{-10}$ mbar), containing a silicate-coated copper sample surface, operating at temperatures between 18 and 400~K. The system is equipped with a quadrupole mass spectrometer (QMS), which is used for the TPD experiments. A sample of either O$_2$, CO or CO$_2$ was deposited onto the surface at 18~K via the triply differentially pumped beam line; a linear temperature ramp was then applied to the surface, and the QMS used to measure the desorption of each species into the gas phase, as a function of temperature.

\begin{table}
  \caption{Description of experimental exposures.}
  \label{tbl:depot}
  \begin{tabular}{@{}lll@{}}
    \hline
    Molecule                        & Surface              & Depositions$^a$\\
                                          &                          & ML     \\ 
    \hline
    \multirow{4}{*}{O$_2$}   & H$_2$O$_{(np)}$ & 0.25, 0.50, 0.75, 1, 1.25$^c$, 1.38, 1.75, \\
                                          &                          & 2.25$^c$, 2.75\\
                                          & H$_2$O$_{(c)}$   & 0.08, 0.17, 0.50, 0.67, 1, 1.33\\
                                          &SiO$_x$              & 0.20, 0.60, 1, 1.20, 1.60, 1.67\\
    \hline
    \multirow{4}{*}{CO}        & H$_2$O$_{(np)}$ & 0.25, 0.50, 0.75, 0.88, 1, 1.06$^b$, 1.13$^b$, \\
                                          &                          & 1.25$^b$, 1.75$^b$, 2.50\\ 
                                          & H$_2$O$_{(c)}$   & 0.10, 0.20, 0.60, 0.80, 1, 1.20$^b$, 1.60\\
                                          &SiO$_x$              & 0.50, 1, 1.25$^b$, 1.75, 2.25\\
    \hline
    \multirow{3}{*}{CO$_2$} & H$_2$O$_{(np)}$ & 1, 2$^b$, 4$^b$, 6$^b$, 10$^b$, 20\\ 
                                          & H$_2$O$_{(c)}$   & 1, 2, 3, 5$^c$, 10, 20\\
                                          &SiO$_x$              & 0.5, 1, 2, 3, 5, 10\\
    \hline
  \end{tabular}

  $^a$ defined as $\frac{exposure}{N_{mono}}$, where N$_{mono}$ is defined in \S~\ref{sec:expt}.\\
  $^b$ Omitted from model (shown in grey in Figure~\ref{fgr:expt}).\\
  $^c$ Modelled, but omitted from Figure~\ref{fgr:model} for clarity.
\end{table}

Three surfaces were investigated in this study: non-porous ASW (H$_2$O$_{(np)}$), crystalline ice (H$_2$O$_{(c)}$) and amorphous silicate (SiO$_x$). The SiO$_x$ surface was recently installed in the FORMOLISM experiment \citep{Lemaire10}, and mimics bare dust grains in molecular clouds. The silicate is amorphous in nature, as evidenced by infrared spectroscopic studies, while TPD experiments, including those presented in this study, reveal the surface to be non-porous on the molecular scale.

The silicate surface is capable of reaching base temperatures of 18~K, which is how the molecular deposition temperature on all the surfaces was pre-determined. For the ice surfaces, 50~ML films were grown on top of the silicate surface by spraying water vapour from a microchannel array doser located 2~cm in front of the surface. The water vapour was obtained from deionised water which had been purified by several freeze-pump-thaw cycles, carried out under vacuum. H$_2$O$_{(np)}$ mimics the ASW which comprises the bulk of interstellar ice, and H$_2$O$_{(c)}$ mimics the crystalline ice seen in some star-forming regions. To produce H$_2$O$_{(np)}$, water was dosed while the surface was held at a constant temperature of 120~K. To form H$_2$O$_{(c)}$, the surface was held at 120~K during the deposition, then flash heated at 50~Kmin$^{-1}$ to 140~K, and finally at 10~Kmin$^{-1}$ to 142.5~K. For each type of ice surface, the temperature was then held constant until the background pressure in the chamber stabilised, before cooling it back down to 18~K, at which temperature adsorbates were dosed onto the respective surfaces.

\begin{figure*}
  \centering
  \includegraphics[width=\textwidth]{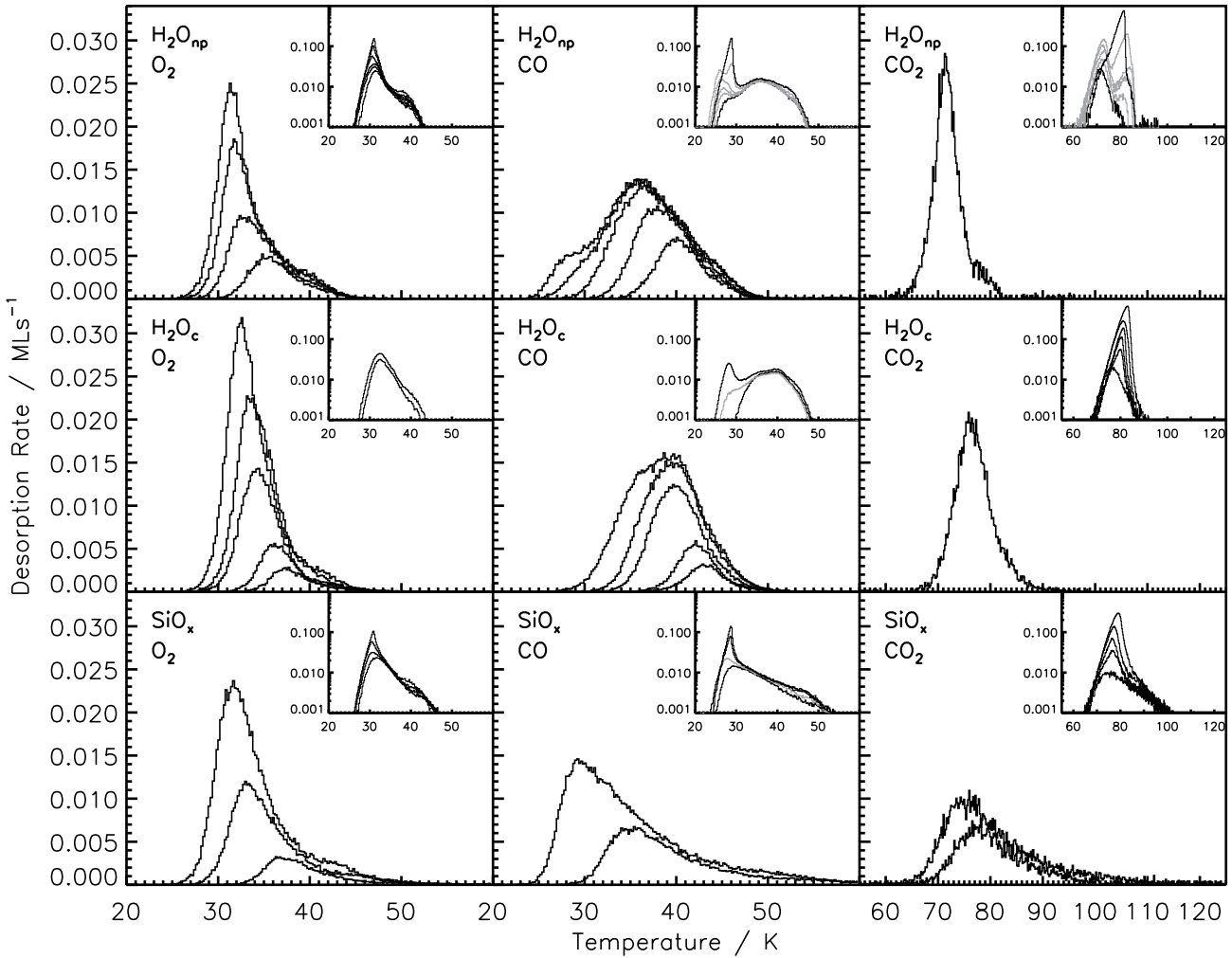}
  \caption{Temperature-programmed desorption spectra of O$_2$, CO and CO$_2$ from H$_2$O$_{(np)}$, H$_2$O$_{(c)}$ and SiO$_x$. For each molecule-surface combination, the main window shows a series of TPD spectra from the lowest sub-monolayer coverage up to 1~ML; inset are TPD spectra from 1~ML up to the highest multilayer coverage, as specified in Table~\ref{tbl:depot}. Spectra shown in black were included in the model, whilst those in grey were omitted. In all cases, the line profile of the TPD spectra in the multilayer regime differ from those in the sub-monolayer regime and, with the exception of CO, the form of the TPD profiles is very similar in the sub-monolayer regime, independent of surface. The desorption profiles of all three molecules from the SiO$_2$ surface in the sub-monolayer regime strongly resemble each other, unlike the ices. See text for full discussion.}
  \label{fgr:expt}
\end{figure*}

For each of the nine combinations of molecule and surface, a series of depositions were made, varying the adsorbate surface coverage, from sub-monolayer to multilayer, as outlined in Table~\ref{tbl:depot}. After the first adsorbate exposure, the surface was then heated at a rate of 10~Kmin$^{-1}$ until the adsorbate was fully desorbed from the surface (below 100~K on the ice surfaces and 130~K on the silicate surface). The original surface therefore remained intact, and by purposefully keeping the temperature of the ice below 100~K no further thermal annealing of the ice occurred during an experimental sequence. Subsequently, the surface was then re-cooled to 18~K, and a new coverage of the same adsorbate added. In this way we could be certain the same surface was used for a single series of adsorbate exposures, allowing real comparison of the coverage effect for the first time. In the multilayer, the underlying surface is generally not important to the desorption, as adsorbate-adsorbate interactions between molecules determine the kinetics. In the sub-monolayer, the interaction between the adsorbate and the substrate is fundamental in determining the desorption characteristics. Thus, by including both regimes, we hoped to probe the differences between these two coverages for different molecules and different surfaces.

A porous ASW surface was not included in this study for two reasons. First, the presence of pores in the ice surface can produce very complex desorption profiles, due to the competing kinetics which arise after entrapment of adsorbates in the pores \citep[e.g.][]{Collings03a}. Worse still, pore collapse starts at only 10--15~K, so the experimental approach used here, of exposing a single ice surface to a variety of adsorbate coverages, is simply not possible with the porous ASW. A new ASW sample would have been required for each experiment; each time we form a porous ice in the laboratory its structure will be slightly different, resulting in a massive change in surface area and non-reproducible adsorbate exposures being required to obtain monolayer coverages. Furthermore, one aim of this experiment is to compare surface characteristics using a single model for the whole dataset, and such an approach would not have been possible when including both porous and non-porous ices, again because the monolayer surface area is difficult to define. Secondly, it has recently been shown experimentally that it is very likely that all water ice in interstellar regions will be non-porous ASW \citep{Accolla11}, even though observations previously suggested that invoking a porous ASW was the only way to explain observed quantities of mixed H$_2$O and CO ices \citep{Pontoppidan03}. 

\section{Results and discussion}

\subsection{Experimental data}\label{sec:expt}

All the TPD results are shown in Figure~\ref{fgr:expt}. The results are ordered top to bottom by surface type, H$_2$O$_{(np)}$, H$_2$O$_{(c)}$ and SiO$_x$, and left to right by adsorbate molecule, O$_2$, CO and CO$_2$. For each molecule-surface combination, the main window shows the series of sub-monolayer exposure TPD spectra, derived from the lowest molecular coverage, up to 1~ML (as per the values given in Table~\ref{tbl:depot}). The inset shows the equivalent data for coverages of 1~ML and above. 

The spectrum which represents 1~ML is chosen by visual inspection of all TPD spectra in the series, and is defined in this study as the highest coverage spectrum which includes only (sub-)monolayer desorption characteristics, and has no multilayer component. For a fully wetting molecule, such as CO or O$_2$, this coverage is assumed to contain 10$^{15}$ molecules cm$^{-2}$, the generally accepted definition of a monolayer \citep{Amiaud2006}. In the case of CO$_2$, however, a lower exposure of the species was required before intermediate and multilayer desorption characteristics were observed i.e. the 1~ML coverage contains fewer than 10$^{15}$ molecules. Thus, in this study, we have defined the coverage as $\theta$~=~$\frac{exposure}{N_{mono}}$, where exposure is the number of molecules deposited on the surface and N$_{mono}$ is the number of molecules in the designated 1~ML deposition. This definition is necessary to allow modelling of the CO$_2$ data in the same fashion as the CO and O$_2$ data, as discussed in \S~\ref{modelling}. 

By inspection of Figure~\ref{fgr:expt} it is immediately apparent that, for all cases, the TPD line profile differs between the sub-monolayer and the multilayer regime, as expected. In some cases (shown in grey on Figure~\ref{fgr:expt}), an intermediate region is visible in those TPD spectra which trace coverages between 1~ML and the multilayer, for example in CO or CO$_2$ on H$_2$O$_{(np)}$. It is clear that structural reorganisations must occur at these intermediate coverages, as desorption spectra with double and triple peaked profiles are observed. Describing such processes is beyond the scope of this paper and for this reason these particular data have been omitted from the later modelling, only being shown here for completeness.

For both O$_2$ and CO$_2$ it is also apparent from Figure~\ref{fgr:expt} that the TPD line profiles are similar in the sub-monolayer regime, independent of the underlying surface. For CO this is clearly not the case, although the multilayer data are more consistent across all surfaces. Likewise, in both the sub-monolayer and multilayer regimes, the TPD spectra for different adsorbates on the SiO$_x$ surface strongly resemble each other; the same can not be said for the ice surfaces. For all the adsorbates, it is evident that multilayer desorption from the crystalline ice surface starts at a slightly higher temperatures than from either the amorphous ice or amorphous SiO$_2$ surface, suggesting a trend dependent on the morphology of the underlying surface. Specifically, for O$_2$, the peak temperature occurs between 30.9 and 32.6~K, but starts around 24--25~K on amorphous surfaces (H$_2$O$_{(np)}$ and SiO$_x$),  compared to 25--26~K from H$_2$O$_{(c)}$.  For CO, the peak temperatures are in the range 28.1--28.9~K, desorption starts at 22--23~K from the amorphous substrates and 23--24~K from H$_2$O$_{(c)}$. For CO$_2$, the peak desorption range is wider (79.3--83.0~K) and desorption begins at 63--65~K from SiO$_x$,  64--65~K  from H$_2$O$_{(np)}$ and 66--67~K from H$_2$O$_{(c)}$. 

\begin{figure}
  \centering
  \includegraphics[width=\columnwidth]{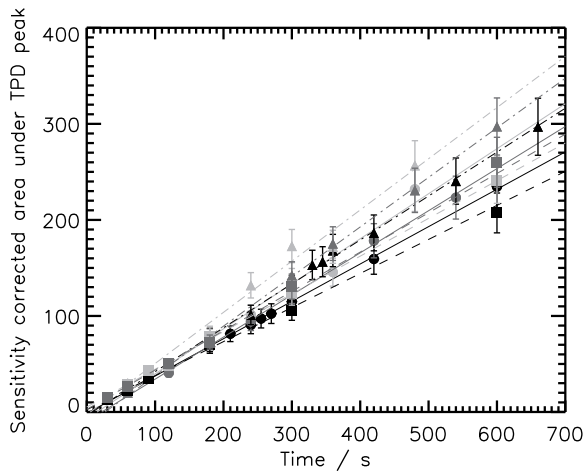}
  \caption{Area under TPD peak, corrected for QMS sensitivity, plotted against deposition time. Experimental data is plotted as points, with H$_2$O$_{(np)}$ in black, H$_2$O$_{(c)}$ in light grey and SiO$_x$ in dark grey; CO as circles, CO$_2$ as squares and O$_2$ as triangles. Overplotted are the straight line fits to the data, as described in the text. Colours are as above for the three surfaces, with CO solid, CO$_2$ dashed and O$_2$ dot-dashed lines. As expected, for each molecule-surface combination, the area increases with deposition time, and the straight line fit goes through zero, within experimental errors. For each molecular species, sticking occurs approximately equivalently on all three surfaces, with an error of $<$~10~\%, although sticking is slightly lower on H$_2$O$_{(np)}$ for all molecules.}
  \label{fgr:sticking}
\end{figure}

To compare the sticking efficiency of each molecule on each surface, the area under the TPD peak, corrected for QMS sensitivity \citep{Matar10}, was plotted against deposition time for each molecule on each surface, see Figure~\ref{fgr:sticking}. Data can be divided into adsorbates -- circle (CO), triangle (O$_2$), and square (CO$_2$) -- and surfaces -- H$_2$O$_{(np)}$ in black, H$_2$O$_{(c)}$ in light grey and SiO$_x$ in dark grey. Such a direct comparison is possible because the experimental method employs the same surface for a series of adsorbate exposures. Furthermore, the dosing rates of all three molecules, and the effective beam pressures employed, are always approximately equal. Straight lines of the form $y = mx + c$ were fitted to each molecule-surface combination and overplotted on the experimental data in Figure~\ref{fgr:sticking}. Within experimental uncertainty (shown as error bars in Figure~\ref{fgr:sticking}), these straight line fits all pass through zero, as expected, indicating that the sticking probability is constant (within $<$~10\%) across the dynamic range of each experiment. This means that no discernible change in sticking probability occurs for any molecule on any surface between the sub-monolayer and multilayer regimes. Adsorbate coverage is always linear as a function of exposure time. Crucially, for all three adsorbate molecules, the sticking efficiency on H$_2$O$_{(np)}$ (black lines in Figure~\ref{fgr:sticking}) is slightly lower when compared with the other two surfaces. This is likely due to fact that the average adsorption energy on H$_2$O$_{(np)}$ peaks at a lower energy than for the other substrates. 

One final point to emphasise is the distinction between the time at which a surface will have been exposed to an equivalent dose of 10$^{15}$ molecules (generally defined as the number of molecules in one monolayer for a wetting molecule), and the coverage (exposure time) at which the TPD data show a deviation from sub-monolayer to multilayer behaviour, defined above as $N_{mono}$. For CO and O$_2$, $N_{mono}$ is reached on all surfaces between around 250--300~s exposure; for CO$_2$ the switch from monolayer to multilayer occurs at much lower surface coverages (exposure times $\sim$~30~\% those for CO and O$_2$). We postulate that this difference is related to the wetting behaviour of the molecules, and come back to this issue in \S~\ref{discussion}.

%
\subsection{Modelling}\label{modelling}

To obtain a more quantitative understanding of the interactions between the molecules and the surfaces, the TPD spectra were reproduced from an empirical kinetic model. To do this, the results were divided into sets of TPD spectra associated with sub-monolayer and multilayer behaviour and the kinetic parameters in each regime determined independently, before being recombined to reproduce the experimental TPD data. A simple model was therefore derived to fit all nine molecule-surface combinations from sub-monolayer to multilayer adsorbate coverages.  

On any surface, at any coverage, thermal desorption can be described in terms of an Arrhenius law, expressed by the Polanyi-Wigner equation, where the rate of desorption, $r$, is described by:
\begin{equation}\label{eqn:polanyi}
  r = -\frac{dN}{dt} = AN^ne^{-E_{ads}/kT},
\end{equation}
where $A$ is the pre-exponential factor, $N$ is the number of adsorbed molecules on the surface (cm$^{-2}$), $n$ is the order of the reaction, $E_{ads}$ is the energy of adsorption of a molecule to the surface (eV), $k$ is the Boltzmann constant (eV molec$^{-1}$K$^{-1}$), and $T$ is the temperature of the surface (K). The units of $A$ depend on $n$: molecules$^{1-n}$cm$^{-2+2n}$s$^{-1}$. Reformulating Equation~\ref{eqn:polanyi} to reflect the measured experimental TPD signal gives:
\begin{equation}\label{eqn:polanyi2}
  r = -\frac{dN}{dT} = \frac{A}{\beta}N^ne^{-E_{ads}/kT},
\end{equation}
where $\beta$ is the rate of heating, in this case $dT/dt$ = 10~Kmin$^{-1}$. 

When the rate of desorption is independent of surface coverage, as is the case in multilayer desorption, $n$~=~0 and the TPD desorption profiles share a common leading edge as is readily apparent from the insets in Figure~\ref{fgr:expt}. 

For each molecule-surface combination, the leading edge of the highest multilayer exposure TPD spectrum was used to determine $E_{ads}$ (in ~K). Although some multilayer desorption TPD data show that the order of reaction can deviate slightly from zero (e.g. \citet{Brown07}) it is a reasonable approximation, so henceforth we assume $n$~=~0. As applied previously by \citet{Acharyya07}, the pre-exponential factor in Equation~\ref{eqn:polanyi2} was assumed to be a function of $E_{ads}$ approximated by:
\begin{equation}\label{eqn:harmonic}
  A = N_{ML}\cdot\nu = N_{ML}\sqrt{\frac{2N_{ML}E_{ads}}{\pi^2M}},
\end{equation}
where $M$ is the mass of the adsorbate molecule, and $N_{ML}$~$\sim$~10$^{15}$~cm$^{-2}$. The key advantage of this method is that the multilayer fitting requires only one variable, $E_{ads}$, removing the interdependency of fitting both the pre-exponential factor and $E_{ads}$ concurrently. The results of fitting $E_{ads}$ are presented in Table~\ref{tbl:multi}.

\begin{table}
  \caption{Multilayer desorption parameters calculated using Equation~\ref{eqn:polanyi2}, compared to previous literature values in italics}
  \label{tbl:multi}
  \begin{tabular}{@{}llll@{}}
    \hline
    Molecule & Surface   & \multicolumn{2}{l}{Calculated values$^a$} \\                             
                   &               & $\nu$                                                 & $E_{ads}$                     \\ 
                   &               & 10$^{26}$~molec~cm$^{-2}$s$^{-1}$   &           K                       \\
    \hline
    \multirow{4}{*}{O$_2$}   & H$_2$O$_{(np)}$ & 6.9 & 898(30) \\
                                          & H$_2$O$_{(c)}$   & 7.0 & 936(40)\\ 
                                          & SiO$_x$             & 6.9 & 895(36)\\ 
                                          & \emph{Au}             & \emph{6.9}  & \emph{912(15)}$^b$\\
                                         &                           & ... & \emph{925(25)}$^c$\\
    \hline
    \multirow{6}{*}{CO}        & H$_2$O$_{(np)}$ & 7.1 & 828(28) \\ 
                                          & H$_2$O$_{(c)}$   & 7.1 & 849(55)\\
                                          & SiO$_x$              & 7.1 & 831(40)\\ 
                                          & \emph{Au}                      & \emph{7.2}  & \emph{858(15)}$^b$\\
                                          &                            & ...    & \emph{826(24)}$^d$\\                                          
                                          &                             & ...    & \emph{855(25)}$^e$\\
    \hline
    \multirow{5}{*}{CO$_2$} & H$_2$O$_{(np)}$     & 9.3 & 2267(71) \\
                                          & H$_2$O$_{(c)}$       & 9.5 & 2356(83)\\ 
                                          & SiO$_x$                & 9.3 & 2269(80)\\ 
                                          & \emph{porous ASW}         & ... & \emph{2690(50)}$^{f}$\\
                                          & \emph{HOPG}         & ... & \emph{2982(-)}$^{g}$\\
    \hline 
  \end{tabular}
$^a$ For values calculated in this work, the values in parentheses are the total calculated error, including the 3~$\sigma$ statistical errors on $E_{ads}$ calculated during the fit of Equation~\ref{eqn:polanyi2} and the experimental uncertainties. In all other cases the error is as quoted in the relevant literature.\\
$^b$ \citet{Acharyya07}, using an identical multilayer modelling method. \\
$^c$ \citet{Fuchs06}, calculated assuming a fixed pre-exponential factor.\\ 
$^{d}$\citet{Collings03b}, calculated assuming a fixed pre-exponential factor.\\
$^{e}$\citet{Oberg05, Fuchs06, Bisschop06}, calculated assuming a fixed pre-exponential factor.\\
$^{f}$ \citet{Sandford90b}, calculated using spectroscopic data, assuming a zeroth order desorption profile.\\
$^{g}$ \citet{Burke10}, calculated empirically with a non-integer reaction order.\\
\end{table}

In the sub-monolayer regime, the rate of desorption is dependent on surface coverage, $\theta = N/N_{mono} \le 1$, where $N_{mono}$ (as defined in \S~\ref{sec:expt}) is the maximum number of molecules on the surface prior to the onset of intermediate or multilayer desorption behaviour. Previous publications have shown that the reaction order can deviate slightly from 1 (e.g. \citet{Brown07}), but for the purposes of this work we assume $n$~=~1. In general, in the case of first order desorption, the peaks of the TPD data appear at a single temperature value, provided that the desorption energy is independent of coverage \citep{Fraser01}. However, as is evident from the main panels in Figure~\ref{fgr:expt}, in this instance it is the trailing edge of the TPD curves that are common in each molecule-surface system. Such behaviour is known to arise when the adsorbate is occupying multiple sites on the surface, such that the adsorption energy is a continuous function of the number of molecules on the surface \citep[e.g.][]{Dohnalek2001,Amiaud2006}. By inverting Equation~\ref{eqn:polanyi2}, a TPD spectrum can be converted to a function, $E(N)$, given by:
\begin{equation}\label{eqn:polanyi3}
  E(N) = -kTln\frac{r\beta}{AN}.
\end{equation}

To fit $E(N)$ for each molecule-surface system, we first select the TPD spectrum equating to the surface exposure just prior to the appearance of multilayer peaks, or (in the case of CO$_2$) more complex desorption characteristics, and define this as the monolayer coverage, i.e. $\theta = N/N_{mono} = 1$, ensuring the maximum range of adsorption sites and energies at the surface are included. This is a very reasonable approach, given that it is well known that $E(N)$ is independent of reaction order at exactly $N =N_{mono}$ \citep{Ulbricht06}.  This monolayer TPD is then fitted using an analytical expression of $E(N)$, redefined in terms of surface coverage, $E(\theta)$, given by:
\begin{equation}\label{eqn:energy}
\begin{split} 
E(\theta) = \alpha_{0} + \alpha_{1}~\theta + \alpha_{2}~\theta^2 + \alpha_{3}~e^{\alpha_{4}~\theta} +\\
\alpha_{5}~e^{\alpha_{6}~\theta} + \alpha_{7}~e^{\alpha_{8}~\theta} + \alpha_{9}~e^{\alpha_{10}~\theta}.
\end{split}
\end{equation}

The resulting coefficients $\alpha_0$~--~$\alpha_{10}$ are listed in Table~\ref{tbl:mono} and can then be used to calculate $E(\theta)$ from Equation~\ref{eqn:energy}, which can then be substituted into Equation~\ref{eqn:polanyi3}, along with an appropriate value for $A$, to reproduce all TPD data for a specific molecule-surface system.

As the choice of $A$ has a large effect on the value of $E(\theta)$ calculated from the model, we first attempted to reproduce the sub-monolayer coverages by concurrently optimising $A$ \citep{Stirniman96}. During the calculation of $E(N)$, $A$ was set at a fixed range of values from 10$^5$--~10$^{15}$~s$^{-1}$. The optimal value of $A$ is that which minimises the difference between the functions $E(N)$ for each sub-monolayer coverage in terms of least-squares. This method was not successful for all molecule-surface combinations, as there was not always a minimum in the least-squares calculation. In those cases where a minimum was calculated, the corresponding $A$ value was found to be significantly lower than the expected value (typically around 10$^6$~s$^{-1}$, whereas the actual $A$ expected was 10$^{12}$). This optimisation method for $A$ assumes that there is no difference in the occupation of the high energy binding sites (i.e. the tail of the TPD spectra) with increasing coverage. Thus, all $E(N)$ calculated for a given $A$ value should coincide at low coverage; in our experiments, there were some minor differences in the tails of the TPD spectra, and consequently this method was rejected. To reproduce the TPD spectra at sub-monolayer coverages $A$ was fixed at 10$^{12}$~s$^{-1}$. 

\begin{table*}
  \centering
  \begin{minipage}{180mm}
  \caption{Calculated coefficients to describe the $E(\theta)$, as in Equation~\ref{eqn:energy}.} 
  \label{tbl:mono}
  \begin{tabular}{@{}lllllllllllll@{}}
    \hline
    Molecule & Surface & \multicolumn{11}{l}{Coefficients} \\
                   &             & $\alpha_{0}$  & $\alpha_{1}$  & $\alpha_{2}$  & $\alpha_{3}$  & $\alpha_{4}$  & $\alpha_{5}$  & $\alpha_{6}$  & $\alpha_{7}$  & $\alpha_{8}$  & $\alpha_{9}$  & $\alpha_{10}$\\
    \hline 
    \multirow{3}{*}{O$_2$} & H$_2$O$_{(np)}$ & 0.080 & 0. & 0. & 0.030 & -4.124 & 0.047 & -441.083 & -3.397 x10$^{-07}$ & 8.449 & 0.006 & -77.198 \\ 
    & H$_2$O$_{(c)}$  & 0.099 & -0.032 & 0.017 & 0.043 & -306.959 & -0.098 & -13026.500 & -2.752 x10$^{-07}$ & 8.212 & 0.016 & -17.518 \\ 
    & SiO$_x$          & 0.028 & 0. & 0. & 0.032 & -7.405 & 0.027 & -118.135 & -3.323 x10$^{-09}$ & 9.398 & 0.067 & -0.248  \\ 
    \hline
    \multirow{3}{*}{CO} & H$_2$O$_{(np)}$ &  0.073 & 0. & 0. & 0.054 & -428.915 & 0.042 & -0.988 & -8.068 x10$^{-06}$ & 7.447 & 0.009 & -14.156 \\ 
    & H$_2$O$_{(c)}$  & 0.107 & -0.028 & 0.011 & 0.029 & -342.873 & 0.010 & -26.879 & -4.150 x10$^{-07}$ & 9.984 & 0.010 & -0.585  \\ 
    & SiO$_x$          & 0.057 & 0. & 0. & 0.060 & -1.176 & 0.034 & -10.199 & -2.716 x10$^{-08}$ & 9.571 & 0. & 0. \\ 
    \hline
    \multirow{3}{*}{CO$_2$} & H$_2$O$_{(np)}$ & 0.195 & -0.015 & 0.007 & 0.117 & -87.247 & -0.082 & -46.774 & 5.035 x10$^{-07}$ & 9.344 & 0.037 & 13.799 \\
    & H$_2$O$_{(c)}$  & 0.201 & 0. & 0. & -32.300 & -6865.170 & 0.026 & -73.854 & 5.640 x10$^{-10}$ & 15.280 & 0.024 & -4.143  \\
    & SiO$_x$          & 0.199 & 0. & 0. & 0.082 & -3.283 & -0.045 & -32.504 & -1.640 x10$^{-05}$ & 5.937 & 0. & 0. \\ 
    \hline
  \end{tabular}
  \end{minipage}
\end{table*}

To successfully reproduce all the TPD data, from sub-monolayer to multilayer regimes, the results from the two coverage models are then combined, such that (a) when $\theta \le 1$, in Equation~\ref{eqn:polanyi2}, $E_{ads}$ is replaced by $E(N)$ (derived from Equation~\ref{eqn:energy}), $n$~=~1, and $N \le N_{mono}$ and (b) when $\theta > 1$, the TPD is the sum of the monolayer trace generated in (a), plus Equation~\ref{eqn:polanyi2}, where $E_{ads}$ is taken from Table~\ref{tbl:multi}, A defined by Equation~\ref{eqn:harmonic}, $n$~=~0 and $N$ is redefined as $N - N_{mono}$.

\subsection{Discussion}\label{discussion}

\begin{figure*}
  \centering
  \includegraphics[width=\textwidth]{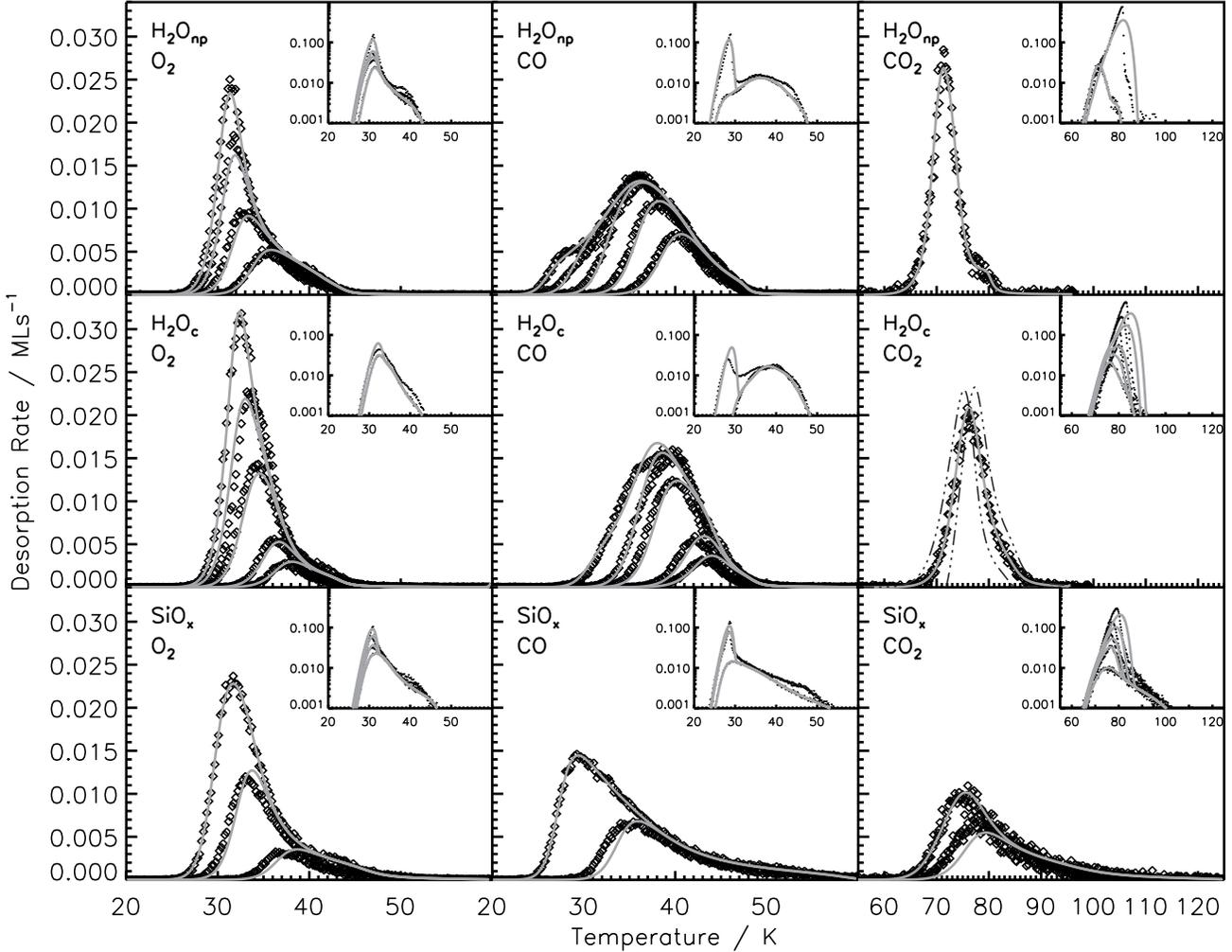}
  \caption{Modelled temperature-programmed desorption spectra. Experimental data, as in Figure~\ref{fgr:expt} are plotted in black diamonds; coverages up to and including 1~ML are shown in the main window, with coverages of 1~ML and greater inset. Model data is overplotted as coloured/grey lines. Typical 3$\sigma$ errors on the model are overplotted as dot-dashed lines on the plot of CO$_2$ on H$_2$O$_{(c)}$ only, for clarity. Some of the original TPD data are omitted for clarity (see Table~\ref{tbl:depot}). See text for full discussion.}
  \label{fgr:model}
\end{figure*}

From Figure~\ref{fgr:model}, it is clear that the combined model can very accurately reproduce the desorption profiles of all molecular species on all surfaces (O$_2$, CO and CO$_2$ on H$_2$O$_{(np)}$, H$_2$O$_{(c)}$ and SiO$_x$). In particular, the model faithfully reproduces the sub-monolayer coverage TPD spectra. Although in a few cases the model data do not perfectly describe the empirical line profiles, the two clearly coincide within the 3~$\sigma$ errors on the model, as illustrated by the overplotted dot-dashed lines on the sub-monolayer coverage of CO$_2$ on H$_2$O$_{(c)}$. These errors are omitted from all the other plots for clarity.  The errors result from the assumption that all the TPD spectra are concurrent in the highest energy binding sites, i.e. their tails overlap, which is a necessary assumption when applying this model.

The multilayer TPD spectra, and their associated fits, are shown in the insets in Figure~\ref{fgr:model}. Most importantly, as noted previously from the experimental data, the model data also show that multilayer desorption begins at the same temperature for each molecule-surface combination (the leading edges of the TPDs are identical), although the onset of this desorption occurs at a higher temperatures on the H$_2$O$_{(c)}$ surface when compared with the amorphous H$_2$O$_{(np)}$ and SiO$_x$ surfaces. From Table~\ref{tbl:multi} it is clear that $E_{ads}$ is consistently marginally greater on the crystalline surface, as compared to the amorphous surfaces, resulting in slightly lower desorption rates from crystalline surfaces, and the later onset (in time or temperature) of the desorption process. This trend is even consistent with previously reported results (see Table~\ref{tbl:multi}). Given that species such as CO, CO$_2$ and $O_2$ could be present on interstellar ices or grains at multilayer coverages of up to a few ML, this result indicates that a similar effect from underlying surface crystallinity could be relevant in interstellar regions. 

It should be noted that the desorption energies of species from H$_2$O$_{(c)}$ are consistently higher than for the same species desorbing from either of the amorphous surfaces, H$_2$O$_{(np)}$ and SiO$_x$. One possible explanation for this trend is that the ordered structure of the crystalline water surface induces larger scale interactions among the desorbed molecules. The study of \citep{Dohnalek99} revealed that amorphous water deposited on top of crystalline water displayed a degree of crystalline nature in its first layer due to interaction with the ordered surface of the underlying crystalline substrate. It is possible that O$_2$, CO and CO$_2$ experience similar interactions with the crystalline water surface, although these bonds would likely be of lower energy than for a water-water system. Another possibility is that, rather than the crystalline surface causing unusual desorption energies, it is in fact the values from amorphous surfaces that are altered due to the presence of defects in the amorphous structure which prevent normal monolayer and multilayer regimes to build up on these surfaces. 

Surprisingly, given its ubiquity in interstellar ices, the multilayer desorption energy of pure CO$_2$ ice is not widely reported in the literature. One value has been measured previously from multilayers of CO$_2$ on HOPG \citep{Burke10}, where analysis proved that the reaction order is zero, and estimated the magnitude of the pre-exponential factor at around 10$^{30}$. However, consistent with the rather complex TPD spectra shown in grey in the insets on Figure~\ref{fgr:expt}, it was very difficult to undertake a detailed analysis of multilayer CO$_2$ desorption from the HOPG surface. Consequently, the value of E$_{ads}$ quoted by \citet{Burke10} is not only somewhat higher than the value calculated here, but also has a quoted error greater than the value of E$_{ads}$ itself ($\pm$~19484~K , which is omitted from Table~\ref{tbl:multi} for clarity). Three other multilayer CO$_2$ desorption measurements are claimed in the literature, as indicated in Table~\ref{tbl:multi}; unfortunately all three were determined from transmission infrared data, assuming the desorption kinetics were first and not zeroth order. Consequently, although included for completeness, as the values are often quoted in models, the numbers are not comparable to the results quoted here.

For multilayer coverages of CO and O$_2$ , the $E_{ads}$ calculated from this model can be compared directly to previous (multilayer) results in the literature (see Table~\ref{tbl:multi}). Within the errors, our results for CO and O$_2$ on H$_2$O$_{(c)}$ - a crystalline surface - are identical to those obtained previously by \citet{Acharyya07} for O$_2$ and CO on a polycrystalline Au substrate, using the same fitting method. This last point is particularly relevant when comparing the results, as both here and in \citet{Acharyya07} the parameterisation of the fit is reduced to one by expressing the pre-exponential factor in terms of $E_{ads}$. Consequently, our pre-exponential factor also correlates well with that reported by \citet{Acharyya07} previously. Even though the calculated $E_{ads}$ for the amorphous surfaces are slightly lower than the result form the H$_2$O$_{(c)}$ surface, they also are the consistent, within the calculated error, with the values from \citet{Acharyya07}. The other values of $E_{ads}$ listed in Table~\ref{tbl:multi} for CO and O$_2$ on polycrystalline Au \citep{Collings03b, Oberg05, Fuchs06, Bisschop06}, were calculated assuming a fixed pre-exponential factor ($A$) of magnitude around 10$^{30}$. As $A$ and $E_{ads}$ are not mutually exclusive, it is not possible to make a direct comparison between these data and our results; however it is interesting to note that despite a difference of over four orders of magnitude in the value of the pre-exponential factor, the reported values of $E_{ads}$ are all remarkably consistent. The question is only the extent to which these subtle differences in the value of E$_{ads}$ affect the desorption of ice multilayers in interstellar space, and how to account for such subtleties in gas-grain models (see \S~\ref{astro}).

In modelling the sub-monolayer desorption of key interstellar molecules from different surfaces, the aim was to provide the chemical modelling community with a useful equation (Equation~\ref{eqn:energy} and Table~\ref{tbl:mono}) for mathematically describing the desorption characteristics of the surfaces, as well as an indication of the coverage at which a particular adsorbate's behaviour switches from the sub-monolayer to the multilayer regimes (a measure of $N_{mono}$). The excellent agreement between experimental data and the model is also validation of our method, described in \S~\ref{modelling}. The only limitation of this method is the certainty with which we have empirically selected the TPD curve used to model sub-monolayer desorption characteristics. This curve represents the highest coverage at which only monolayer desorption characteristics are seen, rather than, necessarily, a coverage of 10$^{15}$ molecules cm$^{-2}$ (the generally accepted definition of one monolayer). However, it is possible that an exposure time of a few seconds more or less could have been a slightly better model system to represent (sub-)monolayer desorption. This is the limitation introduced by our chosen method of deposition calibration.

\begin{table}
  \caption{Calculated values of $E(\theta)$ at a range of sub-monolayer surface coverage values, calculated using Equation~\ref{eqn:energy} and the coefficients from Table~\ref{tbl:mono}.} 
  \label{tbl:submono}
  \begin{tabular}{@{}lllllll@{}}
    \hline
    Mol.                     & Surface              & \multicolumn{5}{l}{$E(\theta)$~/~K} \\
                                        &                          & $E(0.1)$  & $E(0.2)$  & $E(0.5)$  & $E(0.9)$  & $E(1.0)$  \\
    \hline
    \multirow{3}{*}{O$_2$} & H$_2$O$_{(np)}$ & 1161 & 1082 & 972 & 928 & 914\\ 
                                        & H$_2$O$_{(c)}$  & 1149 & 1092 & 1017 & 975 & 969\\ 
                                        & SiO$_x$            & 1255 & 1146 & 1019 & 945 & 930\\ 
    \hline
    \multirow{3}{*}{CO}      & H$_2$O$_{(np)}$ & 1307 & 1247 & 1135 & 956 & 863\\ 
                                        & H$_2$O$_{(c)}$  & 1330 & 1288 & 1199 & 1086 & 1009\\ 
                                        & SiO$_x$           & 1418 & 1257 & 1045 & 896 & 867\\ 
    \hline
    \multirow{3}{*}{CO$_2$}& H$_2$O$_{(np)}$& 2346 & 2258 & 2197 & 2197 & 2236\\
                                        & H$_2$O$_{(c)}$  & 2514 & 2451 & 2364 & 2341 & 2361\\
                                        & SiO$_x$           & 3008 & 2798 & 2487 & 2317 & 2271\\ 
    \hline
  \end{tabular}
\end{table}

Of all the data fitted here only the CO-ice systems show minute deviations at the highest sub-monolayer coverages, and such effects could be due to CO, as the smallest molecule studied, somehow being able to probe defects and cracks in the ice surface, prior to forming a multilayer, as described previously in studies of CO on ASW \citep{Collings03a,Collings03b}, an effect omitted from this model. This may also explain why in these systems it is difficult to reproduce a perfect match in the transition regime between the monolayer and multilayer coverages. Likewise, although the fits shown here to the CO$_2$-ice systems are excellent, a number of 'intermediate' TPD spectra, shown in grey in Figure~\ref{fgr:expt}, with double and triple TPD peaks, had to be omitted from the modelling process. These spectra are indicative of 2D island growth on a surface, as seen in the formation of N$_2$ ices, and indicates that the growth mechanism of CO$_2$ on any ice surface is different to that of CO or O$_2$ \citep{Oberg05,Fayolle11}. Whilst each island is entirely independent of another on the surface and there is no diffusion between them or bridges linking them, we can consider that we are in the sub-monolayer coverage regime. Once the islands have fully merged and overlayers of CO$_2$ ice form, then we can be certain we are in a multilayer regime. This is what we assume here; again a more full description of the 'intermediate' stages is beyond the scope of this paper. Nevertheless, such effects should not detract from the broad applicability of the sub-monolayer model. Furthermore, as a general rule the outcome of our monolayer TPD choice suggests that for CO and O$_2$ it is reasonable to assume $N_{mono}$~=~10$^{15}$, whereas for CO$_2$ it is around $N_{mono}$~=~3~$\times$~10$^{14}$. This result is consistent with the exposures discussed previously in Figure~\ref{fgr:sticking} and reflects the CO$_2$ island growth.

As can be seen from the main panels in Figure~\ref{fgr:model}, at very low coverages ($<$~0.5~ML) there is sometimes a small deviation in the leading edge of the modelled desorption compared to the experimental results, most evident on the SiO$_{x}$ surface. Although this lies within the error range of the fitting, there are two potential explanations for why such discrepancies could arise, both related to the assumption made in the model that all desorptions from a given surface share the same TPD trailing edge profile. As the sites in the tail of the TPD are the most energetically favourable, it follows that any adsorbate able to diffuse across the surface will bind most readily to these sites, occupying them first on adsorption and last on desorption. Such an effect has clearly been seen previously with H$_2$ and D$_2$ on ice surfaces \citep{Amiaud2006}. Even under ultra-high vacuum conditions, molecules such as H$_2$, CO and H$_2$O are still present in the experimental chamber, all be it at exceptionally low concentrations (partial pressures around 10$^{-13-14}$). Consequently, over the very long timescales required to complete the experiments described here (14--16 hours per single molecule-surface combination), it is potentially possible that some of the surface sites could be dynamically occupied for a short period by a 'pollutant' adsorbate, thus blocking the adsorbate molecule of interest from occupying a highest energy binding site. Such an effect would then lead to higher occupation of the lower energy binding sites and a miss match between the experimental and model data, the latter appearing at slightly higher temperatures. Such an effect would not have been evident in the work of \citet{Amiaud2006} because they were only cyclically heating their ice surfaces between 10 and 30~K, and consequently could complete the experiment on a much shorter timescale than was possible here. As H$_2$ and CO are reasonably volatile pollutants, the build up of H$_2$O would be most likely to have an effect on the nature of the surface, and such an effect would be most evident on the non ice surface, i.e. SiO$_{x}$. However as the experimental sequence was always run from low to high coverage, if this were the key reason for any discrepancy, it should be more pronounced as we get towards the monolayer coverage data. Evidently this is not the case. 

It is also important to consider mass effects; in comparison to the H$_2$-ice system studied previously \citep{Amiaud2006}, CO, O$_2$ and CO$_2$ are more massive molecules, and therefore likely to diffuse more slowly on the surfaces. Therefore, unlike the H$_2$, when deposited on the surface at relatively low concentrations, CO, O$_2$ and CO$_2$ may not sample all the binding sites before adsorbing, just ballistically depositing in a 'stick and stop' process. Subsequently, during the TPD warming, all the molecules will diffuse, but again the heavier molecules may not fully sample the surface before desorbing. Then, at very low coverages, not all the highest energy binding sites will be occupied, and the TPD tails may not be quite coincident. As a result, the model would slightly underestimate the leading edge of such TPD curves. From this explanation it is also possible to rationalise why the discrepancy between the model and empirical data would be greatest on the SiO$_x$ surface: the breadth of the sub-monolayer TPD peaks from the molecule-silicate systems in comparison to the molecule-ice systems clearly indicates a broader range of binding sites on the SiO$_x$ surface. By contrast, the water surfaces have a narrow, and an energetically similar range of binding sites, which may also intimate that the ice surfaces do not have many dangling bonds or much proton disorder in them \citep{Fraser04}. This is contrary to what we might expect, as it is often assumed that amorphous water ice potentially has a very broad range of binding sites at its surface. However without spectroscopic data we cannot comment further on this here. Consequently, the probability of sampling the highest energy binding sites is even lower at low coverages on SiO$_x$ than the ice surfaces, provided the diffusion rates on all three surfaces are approximately equal. Nevertheless, these tiny differences should not detract from the excellence of the fits, nor that the model can be executed whilst ignoring the pumping speed in the chamber; this is because in FORMOLISM the pumping is so effective that as soon as a molecule desorbs from the surface it is effectively removed from the chamber - in experiments without such efficient pumping the experiments described here would simply be impossible to undertake, as the TPD tails may never overlap. 

To compare the $E(\theta)$ calculated here with previous adsorption energy values of CO, O$_2$ and CO$_2$ as reported in the literature, we have produced Table~\ref{tbl:submono}. This shows the specific values of $E(\theta)$ for each molecule on each surface at coverages ranging from $\theta$~=~0.1--1.0. The errors on each energy value are determined by the error on the fit to the experimental data and are around $\pm$~25~K in all cases. The comparison is rather complicated by the fact that no-one else has ever considered fitting coverage dependent adsorption energies for these particular molecules at sub-monolayer coverages on surfaces of astrophysical relevance. In general the effective adsorption energy is higher at lower coverage, reflecting that the highest energy binding sites are occupied first, and tends towards the calculated multilayer value for the same molecule-surface system as the coverage rises towards that of a monolayer. Although, by the monolayer coverages, the trend again emerges that the binding energy on the crystalline surface is consistently higher than on the amorphous surfaces; the same is not true at lower coverages. 

Previous studies using different methods have reported the monolayer energy of CO as: 1564~$\pm$~120~K on HOPG \citep{Ulbricht06}, 1624~$\pm$~360~K on meteorite \citep{Mautner06}, and 1179~$\pm$~24~K on a highly porous ASW \citep{Collings03b}. Certainly our values for CO on the ice surfaces cover a range of values including the result of \citet{Collings03b}, and at the lowest coverages our value on the SiO$_2$ surface is closer to that reported for HOPG. it is pertinent to compare the lowest coverage value since the HOPG data were measured in a molecular beam scattering experiment, indicating that only a low concentration, dynamic coverage was resident on the HOPG surface for a relatively short time, and therefore must have been occupying the highest energy binding sites. The meteorite sample would no-doubt encompass a much wider variety of binding sites than event the amorphous silicate used here, so again it is not surprising that the desorption energy is even higher on this surface. This excellent agreement between our and previous results is further justification that our model is viable and effectively reproduces the effects of sub-monolayer desorption from a variety of interstellar-relevant surfaces. 

For O$_2$, previous calculated sub-monolayer energies are: 1082~$\pm$~120~K on HOPG \citep{Ulbricht06}, 1203~K on graphite \citep{Bojan87}. Again our values at the lowest coverages on SiO$_2$ compare favourably with previously reported values from carbon-based surfaces. The desorption energy of the CO$_2$ monolayer has been calculated as: 2766~$\pm$~241~K on HOPG \citep{Ulbricht06}, 3079~$\pm$~20~K on graphite \citep{Terlain83}, 2553~$\pm$~232~K on H$_2$O$_{(c)}$ \citep{Andersson04}, 2490~$\pm$~240~K on H$_2$O$_{(np)}$ and 2393~$\pm$~240~K on H$_2$O$_{(c)}$ \citep{Galvez07}. Certainly the range of our data compare favourably with these values; in particular at the lowest coverages, the value of $E(\theta)$ for CO$_2$ on SiO$_2$ is consistent with the molecular beam scattering result from HOPG, and the value of  $E(\theta)$ for CO$_2$ on  H$_2$O$_{(c)}$ is identical within the quoted errors to that measured by \citet{Andersson04}, who also used a molecular beam scattering experiment and therefore probed the highest energy binding sites for CO$_2$ on the crystalline ice surface. Interestingly, the value calculated by \citep{Galvez07} for desorption from H$_2$O$_{(np)}$ is higher than that from H$_2$O$_{(c)}$, a result in contrast to the data presented here. However, within the stated experimental errors of the study, these values are consistent with our results. Such excellent overlap between the results also verifies our earlier assumptions in terms of choosing the appropriate TPD spectrum to define as the monolayer coverage for CO$_2$ on the ice surfaces, and indeed our calculation of $N_{mono}$. 

\begin{figure}
  \centering
  \includegraphics[width=\columnwidth]{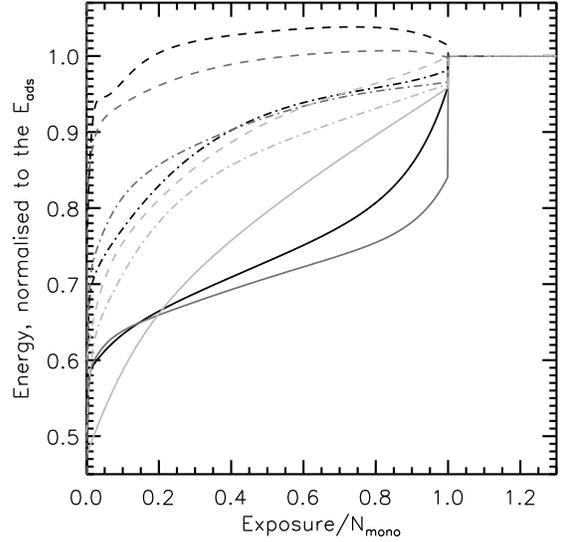}
  \caption{Total modelled energy as a function of exposure/N$_{mono}$ for all molecule-surface combinations. Plots are as follows: O$_2$ dot-dashed, CO solid, CO$_2$ dashed; H$_2$O$_{(np)}$ black, H$_2$O$_{(c)}$ dark grey, SiO$_x$ light grey.}
  \label{fgr:model2}
\end{figure}

The adsorption behaviour of the molecule-surface combinations can be divided into roughly three categories related to the difference between the multilayer energy (or adsorbate-adsorbate interactions) and the energy distribution of the surface adsorption sites. These behaviours can be explained with reference to Figure~\ref{fgr:model2}, a plot of the modelled energy as a function of coverage, $E(\theta)$. Where the energy of the multilayer is significantly higher than the energy distribution at coverages below 1~ML, as is the case for CO on H$_2$O$_{(c)}$, molecules will preferentially adsorb onto the substrate at low coverages, rather than starting to form a multilayer or island. In this 'full wetting' regime, the multilayer appears only after all of the sites on the surface are filled. All the CO lines (solid) in Figure~\ref{fgr:model2} follow this regime.

Where the multilayer energy is lower than the monolayer energy, such as for CO$_2$ on H$_2$O$_{(np)}$, molecules preferentially form clusters (or islands) on the surface at very low coverages, well before 1~ML. In this 'non-wetting' regime, these islands, which are multilayers thick, do not cover the whole substrate but eventually
build long towers which may join and overlap, until eventually multilayer zeroth order desorption is observed. All the CO$_2$ sub-monolayer coverages on any H$_2$O ice surface follow this regime, see Figure~\ref{fgr:model2}.

These two 'wetting' and 'non-wetting' regimes can be further contrasted by considering the TPD spectra of CO on H$_2$O$_{(c)}$ versus those of CO$_2$ on H$_2$O$_{(np)}$ in Figure~\ref{fgr:expt}. CO$_2$ rapidly forms a double-peaked multilayer desorption profile while CO shows a gradual filling of the surface sites before a multilayer feature forms.

Intermediate between these two regimes, molecules first fill the lowest energy surface sites, before starting to form multilayer islands on top of the existing monolayer, before the monolayer coverage is fully complete. This occurs where the energy of the final binding sites in the monolayer is equal to or greater than that
of the multilayer binding sites, and usually is only seen at exposures close to N$_{mono}$. These three states are relatively easy to observe in Figure~\ref{fgr:model2}, although the exact boundaries between the regimes are difficult to define, as island growth is difficult to reproduce and analyse experimentally. Nevertheless, the regimes associated with each molecule-surface combination are summarised in Table~\ref{tbl:behaviour}.

\begin{table}

  \caption{Adsorption behaviour of each molecule-surface combination, estimated from the modelled energies in Tables~\ref{tbl:multi}~\&~\ref{tbl:mono} and the desorption characteristics in Figure~\ref{fgr:expt}.$^a$}
  \label{tbl:behaviour}
  \begin{tabular}{@{}lll@{}}
    \hline
    Molecule & Surface   & Behaviour \\                             
    \hline
    \multirow{3}{*}{O$_2$}   & H$_2$O$_{(np)}$ & Full wetting / Intermediate\\
                                          & H$_2$O$_{(c)}$   & Full wetting / Intermediate\\ 
                                          & SiO$_x$             & Full wetting \\ 
    \hline
    \multirow{3}{*}{CO}        & H$_2$O$_{(np)}$ & Full wetting\\ 
                                          & H$_2$O$_{(c)}$   & Full wetting\\
                                          & SiO$_x$              & Full wetting\\ 
    \hline
    \multirow{3}{*}{CO$_2$} & H$_2$O$_{(np)}$ & Non-wetting\\ 
                                          & H$_2$O$_{(c)}$   & Non-wetting\\ 
                                          & SiO$_x$             & Non-wetting / Intermediate\\ 
    \hline
  \end{tabular}

$^a$ 'Full wetting' behaviour is seen for molecules which fill all the lowest energy sites on the surface before forming a multilayer. 'Intermediate' behaviour is when the molecules start to fill the surface, but form islands before it is full. 'Non-wetting' behaviour is when islands are almost immediately formed by molecules on the surface.\\
\end{table}

\section{Astrophysical Implications}\label{astro}

\begin{figure}
  \centering
  \includegraphics[width=0.5\textwidth]{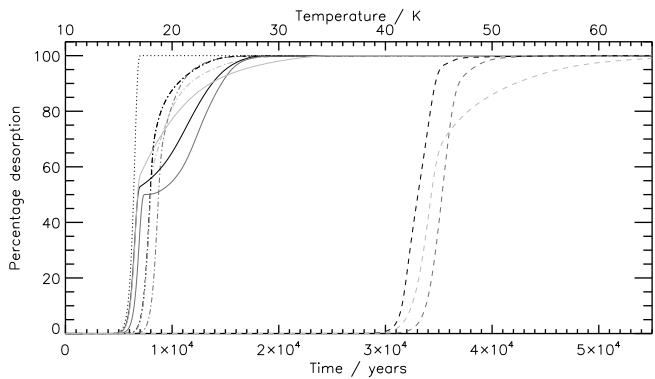}
  \caption{The simulated desorption of 2~ML of CO (solid), O$_2$ (dot-dashed) and CO$_2$ (dashed) from H$_2$O$_{(np)}$ (black), H$_2$O$_{(c)}$ (dark grey) and SiO$_x$ (light grey) at a heating rate of 1~Kcentury$^{-1}$. For comparison, the black dotted line shows the modelled desorption characteristics of CO on H$_2$O$_{(np)}$ if only multilayer desorption is included in the model.}
  \label{fgr:astro}
\end{figure}

Under astrophysical conditions, the heating rate is likely to be on the order of 1~Kcentury$^{-1}$ (hot core heating rate \citep{Viti99}), many orders of magnitude slower than the laboratory value of 10~Kmin$^{-1}$ employed here. By simulating the desorption profiles of the molecule-surface combinations here, it is possible to investigate whether surface type and coverage is likely to affect desorption characteristics on astrophysically relevant timescales.

Figure~\ref{fgr:astro} presents the results of such a simulation for all molecule-surface combinations, assuming a 2~ML coverage of each species on each grain (i.e. one multilayer and one monolayer). This seems reasonable as none of these species (except possibly CO) form large pure ice overlayers in the ISM. In Figure~\ref{fgr:astro}, the dot-dashed line shows the modelled desorption characteristics of 2~ML of CO from H$_2$O$_{(np)}$ if only multilayer desorption is considered. It is evident that without including the sub-monolayer component of the model, the desorption time is vastly underestimated (by over a few $\times$ 10$^4$~years or 10 -- 15~K), particularly for CO and CO$_2$. This is a comparable error to that introduced when gas-grain models treat multilayer desorption as first order, as suggested previously \citep{Fraser01}, and illustrates that sub-monolayer desorption characteristics are fundamental to describing the overall desorption profile of species on astrophysical timescales.

Figure~\ref{fgr:astro} shows that for O$_2$ there is the least difference between the time (or temperature) at which desorption would be completed in the multilayer only model versus the combined multilayer-monolayer model. However, monolayers of both CO and CO$_2$ are able to reside directly at the surface for thousands of years, or at equivalent temperatures 10 -- 15~K higher, than those predicted by multilayer desorption alone, particularly on the amorphous silicate surface. In quiescent regions of the interstellar medium or at the edges of molecular clouds, where grains are warmer, this would suggest that small molecules, like CO and CO$_2$, could easily be present on the grain surfaces, potentially undergoing chemistry to generate larger chemical species, long before detectable ice layers really form. Such processes have been postulated previously \citep{Fraser05}, and it remains an interesting follow-on from this work to see which chemical species will form in astrochemical models when these sub-monolayer desorption data are considered.

What is also clear from Figure~\ref{fgr:astro} is that, while the molecules desorb at different times (and temperatures), there is a clear pattern to the order of desorption from the various substrates. The onset of desorption from the amorphous surfaces, H$_2$O$_{(np)}$ and SiO$_x$, occurs almost concurrently for all species, while
desorption from H$_2$O$_{(c)}$ occurs later, at a higher temperature. This delay in desorption from crystalline surfaces indicates that in regions where crystalline water is predicted to dominate, for example on processed ices inside disks around YSOs \citep{Visser09}, or even cometary surfaces \citep{Roush01}, volatile species will not
start desorbing from grain surfaces for a few thousand years longer (or over temperature ranges of a few kelvin higher) than anticipated from zeroth order desorption kinetics models. Such differences are subtle, but when one considers that these molecules will be highly mobile at such temperatures, the potential for complex chemistry to occur on the grain surfaces is increased. Furthermore, it is exactly these inner disk regions, where crystalline water ice is predicted to be found, from which hot core and hot corino gases are populated by desorbing molecules\citep[e.g.][]{Viti99,vandertak00,Cazaux03,Semenov10,Favre11}.

Similarly, the simulation results in Figure~\ref{fgr:astro} show the residence time of molecules on a H$_2$O$_{(c)}$ surface is significantly higher than on the amorphous surfaces. Starting from a 2~ML coverage, the time taken for 50~\% of the O$_2$ on SiO$_x$ or H$_2$O$_{(np)}$ to desorb is 8000~years, while on H$_2$O$_{(c)}$ this value is 8700~years, an increase of 8.75~\%. Similarly, the difference between desorption of 50~\% of the molecules from H$_2$O$_{(np)}$ and from H$_2$O$_{(c)}$ is 8.82~\% for CO and 6.97~\% for CO$_2$.

\section{Conclusions}

A key aim of this work was to address what effect the underlying dust grain surface could have on the desorption characteristics of molecules adsorbed there, and how this desorption is modified in the sub-monolayer coverage regime compared to the multilayer regime.

This work used temperature programmed desorption to measure the desorption characteristics of O$_2$, CO and CO$_2$ from H$_2$O$_{(np)}$, H$_2$O$_{(c)}$ and SiO$_x$ over sub-monolayer to multilayer coverages. The experimental data were modelled using the Polanyi-Wigner equation, combining different approaches to reproduce
the sub-monolayer and multilayer TPD traces, in particular by varying E$_{ads}$ as a function of coverage in the sub-monolayer regime. The desorption can be categorised as 'full wetting', 'intermediate' or 'non-wetting' behaviour, and the switching point is well described by N$_{mono}$, an empirical measure of the coverage at which pure monolayer desorption stops. On both a laboratory and an astrophysically relevant timescale, the desorption characteristics of molecules from the amorphous substrates
H$_2$O$_{(np)}$ and SiO$_x$ were found to be very similar, while on the crystalline surface H$_2$O$_{(c)}$, molecules desorbed at higher temperatures and on a longer timescale.

It would be relatively trivial to implement the data presented here into astrochemical models; it would require only a subroutine to describe the sub-monolayer, coverage dependent, E$_{ads}$, which can be calculated using the parameters presented in this paper, and the coverage N$_{mono}$, at which the true monolayer desorption ends and beyond which a fixed E$_{ads}$, and zeroth order desorption can be implemented.

Previously, \citet{Green09} concluded that surface type is relatively unimportant, and rather it is the heating rate and grain size that dominate desorption kinetics; we do not disagree that both these factors are important, but by not considering the sub-monolayer case, this previous work overlooks certain intricacies. Firstly, the results presented here suggest that the surface type is the dominant factor controlling desorption under sub-monolayer coverage conditions. This definition of surface type must include both the surface material and the degree of crystallinity, and not simply the material alone. From the data presented here, it is clear that the desorption characteristics of molecular species from H$_2$O$_{(np)}$ and SiO$_x$, two amorphous surfaces, are much more similar than those from H$_2$O$_{(np)}$ and H$_2$O$_{(c)}$, which are composed of the same underlying material. Secondly, while the size of the grain is critical to desorption characteristics, it is both the surface area and the surface-adsorbate interaction which must be considered. Without including information on whether the surface-adsorbate interaction is wetting or non-wetting, it is not possible to know the coverage at which behaviour switches from sub-monolayer (to intermediate) to the multilayer regime, i.e. the value of N$_{mono}$. As it is notoriously difficult to calculate physisorption surface-adsorbate energies theoretically, this is necessarily an empirically determined value, highlighting the importance of experiments such as those presented here.

\section*{Acknowledgments}

J.A.N. gratefully acknowledges the financial support of The Leverhulme Trust, the Scottish International Education Trust, the University of Strathclyde and the Scottish Universities Physics Alliance, without whom this work would not have been carried out. The research leading to these results has received funding from the European Community's Seventh Framework Programme FP7/2007-2013 under grant agreement no. 238258. We acknowledge the support of the national PCMI programme founded by the CNRS, the Conseil Regional d'Ile de France through SESAME programmes (contract I-07-597R), the Conseil G\'{e}n\'{e}ral du Val d'Oise and the Agence Nationale de Recherche (contract ANR 07-BLAN-0129). The authors would like to thank H. Mokrane for her assistance in the laboratory, as well as Louis d'Hendecourt and Zahia Djouadi for preparing the silicate surface used in these experiments. We are also immensely grateful to the anonymous referee, whose insightful comments have greatly improved this manuscript.

\bsp

\label{lastpage}

\end{document}